\documentclass{Rinton-P9x6}

\begin{document}

\title{Quantum Effects during Inflation}

\author{R. P. Woodard}

\address{Department of Physics, University of Florida, Gainesville, 
FL 32611, USA \\ E-mail: woodard@phys.ufl.edu}

\maketitle

\abstracts{The expansion of spacetime alters the energy-time uncertainty
principle, allowing virtual particles to persist longer than in flat 
space. For a given expansion the effect is largest for massless particles
which are not conformally invariant. For a given particle type the effect
is largest for inflation. I exhibit these two principles in the context of 
massless fermions which are Yukawa-coupled to massless scalars. I also 
review the many related examples which have been studied recently.}

\section{Introduction}

The invariant interval of a homogeneous, isotropic and spatially flat
geometry can be written in the form,
\begin{equation}
ds^2 = -dt^2 + a^2(t) d\vec{x} \! \cdot \! d\vec{x} \; .
\end{equation}
Although the scale factor $a(t)$ is not observable, its derivatives can be 
formed into the Hubble parameter and the deceleration parameter,
\begin{equation}
H(t) \equiv \frac{\dot{a}}{a} \qquad , \qquad q(t) \equiv -
\frac{a \ddot{a}}{\dot{a}^2} = -1 - \frac{\dot{H}}{H^2} \; . \label{HI}
\end{equation}
Inflation is defined as a period of accelerated expansion,\cite{Linde}
that is, $H(t) > 0$ and $q(t) < 0$. Of fundamental importance to the theory
of inflation is the conformal time interval,
\begin{equation}
{\Delta \eta}(t) \equiv \int_0^t \!\! \frac{dt'}{a(t')} \; .
\end{equation}
During non-accelerated expansion ${\Delta \eta}(t)$ grows without bound
whereas it is bounded during inflation, even if the universe inflates 
forever. This simple fact is behind the inflationary resolution of the 
smoothness problem.\cite{Linde} We will see that it also explains why 
certain particles experience large quantum effects during inflation.

Parker was the first to report quantitative results about how the expansion 
of spacetime affects virtual particles.\cite{Parker} The physics is simple 
to understand in the context of the energy-time uncertainty principle and 
the fact that the energy of a particle with mass $m$ and co-moving wave 
vector $\vec{k}$ is,
\begin{equation}
E(t,\vec{k}) = \sqrt{m^2 + \Vert \vec{k} \Vert^2/a^2(t)} \; .
\end{equation}
Three conclusions result:\cite{PW1,W}
\begin{enumerate}
\item{Growth of $a(t)$ always increases the time a virtual particle of
fixed $m$ and $\vec{k}$ can exist;}
\item{The persistence time for given $a(t)$ and $\vec{k}$ is longest for 
smallest $m$; and}
\item{Virtual particles with $m=0$ can persist forever during inflation.}
\end{enumerate}
These principles govern what happens once a virtual particle emerges from
the vacuum. Also important is the emergence rate, which depends upon the 
type of particle. Most massless particles are conformally invariant. This
causes the rate at which they emerge from the vacuum to fall like 
$1/a(t)$.\cite{PW1,W} So any particles that emerge can persist forever 
during inflation, but very few emerge. Hence conformally invariant particles
engender no significant quantum effects during inflation.

Two familiar particles are both massless and also not conformally invariant:
gravitons and massless, minimally coupled scalars. This means they can 
mediate large quantum effects during inflation. Starobinski\u{\i} was
the first to compute the contribution of inflationary gravitons to the
cosmic microwave anisotropy.\cite{AAS} This effect may be observable in the
precision measurements of polarization planned for the next decade. Mukhanov
and Chibisov were the first to suggest that fluctuations from inflationary
scalars may have produced the tiny inhomogeneities needed to form the
various cosmic structures of today under the influence of gravitational 
collapse.\cite{MC} The imprint of these fluctuations on the cosmic microwave 
anisotropy has been imaged with stunning precision by WMAP.\cite{WMAP}

A variety of related quantum effects have been studied from 
gravitons\cite{TW1,TW2} and from scalars\cite{AW,OW,PTW1,PTW2,PW2,PW3}.
The most recent of these is the creation of massless fermions during 
inflation. These particles are conformally invariant so they experience 
no substantial particle creation by themselves. However, when Yukawa-coupled 
to a massless, minimally coupled scalar, there is copious production of
fermions from the spontaneous appearance of a scalar and a fermion-anti-fermion
pair.\cite{PW1} In this paper I consider the same process for flat space,
and for a conformally coupled scalar, in order to illustrate the crucial roles
of inflation and conformal non-invariance.

\section{Particle Creation in Yukawa Theory}

Dirac fermions require gamma matrices $\gamma^b$ which anti-commute as
usual, $\{\gamma^b,\gamma^c\} = -2 \eta^{bc} I$. Coupling fermions to 
gravity also requires the vierbein field, $e_{\mu b}(x)$, the contraction 
of two of which gives the metric, $g_{\mu\nu} = e_{\mu b} e_{\nu c} 
\eta^{bc}$. In a general vierbein background the action we wish to study is,
\begin{eqnarray}
\lefteqn{{\cal L} = -\frac12 \partial_{\mu} \phi \partial_{\nu} \phi 
g^{\mu\nu} \sqrt{-g} - \frac12 \xi \phi^2 R \sqrt{-g} } \nonumber \\
& & \hspace{2cm} + \, \overline{\psi} e^{\mu}_{~b} \gamma^b \Bigl( 
\partial_{\mu} - \frac12 A_{\mu c d} J^{cd} \Bigr) \psi \sqrt{-g} - 
f \phi \, \overline{\psi} \psi \sqrt{-g} \; .
\end{eqnarray}
Here the spin connection and the Lorentz representation matrices are,
\begin{equation}
A_{\mu bc} \equiv e^{\nu}_{~b} \Bigl(e_{\nu c,\mu} - \Gamma^{\rho}_{~\mu\nu}
e_{\rho c} \Bigr) \qquad , \qquad J^{bc} \equiv \frac{i}4 [\gamma^b,\gamma^c] 
\; .
\end{equation}
When $\xi = \frac16$ the scalar is conformally coupled; when $\xi = 0$ it is
minimally coupled.

For the special case of a homogeneous and isotropic metric (\ref{HI}) the 
associated vierbein can be taken to be,
\begin{equation}
e_{\mu b}(t,\vec{x}) \Bigl\vert_{\rm HI} = - \delta_{\mu 0} \delta_{b 0}
+ a(t) \delta_{\mu i} \delta_{b i} \; .
\end{equation}
In this case the Lagrangian simplifies dramatically,
\begin{eqnarray}
\lefteqn{{\cal L} \Bigl\vert_{\rm HI} = \frac{a^3}2 \Bigl(\dot{\phi}^2 - 
\frac1{a^2} \vec{\nabla} \phi \cdot \vec{\nabla} \phi - \xi (12 H^2 + 6 
\dot{H}) \phi^2\Bigr) } \nonumber \\
& & \hspace{3cm} + (a^{\frac32} \overline{\psi}) \Bigl(\gamma^0 \partial_0 + 
\frac1{a} \gamma^i \partial_i \Bigl) (a^{\frac32} \psi) - f a^3 \phi 
\, \overline{\psi} \psi \; . \label{LHI}
\end{eqnarray}
The conformal invariance of free fermions implies that (\ref{LHI}) becomes
even simpler when expressed in terms of conformally rescaled fields,
\begin{equation}
\Psi(t,\vec{x}) \equiv a^{\frac32}\!(t) \psi(t,\vec{x}) \qquad , \qquad
\overline{\Psi}(t,\vec{x}) \equiv a^{\frac32}\!(t) \overline{\psi}(t,\vec{x})
\; .
\end{equation}

Because (\ref{LHI}) possesses spatial translation invariance the three free
fields can be expanded in spatial plane waves,
\begin{eqnarray}
\phi_I(t,\vec{x}) & \! = \! & \int \!\! \frac{d^3k}{(2\pi)^3} \left\{
e^{i \vec{k} \cdot \vec{x}} A(t,k) \alpha(\vec{k}) + e^{-i \vec{k} \cdot
\vec{x}} A^*\!(t,k) \alpha^{\dagger}(\vec{k}) \right\} , \\
\Psi_{\! I}(t,\vec{x}) & \! = \! & \int \!\! \frac{d^3q}{(2\pi)^3} \sum_r 
\left\{ e^{i \vec{q} \cdot \vec{x}} B(t,\vec{q},r) \beta(\vec{q},r) + 
e^{-i \vec{q} \cdot \vec{x}} C(t,\vec{q},r) \gamma^{\dagger}(\vec{q},r) 
\right\} , \\
\overline{\Psi}_{\! I}(t,\vec{x}) & \! = \! & \int \!\! \frac{d^3p}{(2\pi)^3} 
\sum_s \left\{e^{i \vec{p} \cdot \vec{x}} \overline{C}(t,\vec{p},s) 
\gamma(\vec{p},s) + e^{-i \vec{p} \cdot \vec{x}} \overline{B}(t,\vec{p},s) 
\beta^{\dagger}(\vec{p},s) \right\} .
\end{eqnarray}
The various creation and annihilation operators are canonically normalized,
\begin{eqnarray}
\Bigl[\alpha(\vec{k}),\alpha^{\dagger}(\vec{k}')\Bigr] & = & (2\pi)^3 
\delta^3(\vec{k} - \vec{k}') \; , \\
\Bigl\{\beta(\vec{q},r),\beta^{\dagger}(\vec{p},s)\Bigr\} & = & \delta_{rs} 
(2\pi)^3 \delta^3(\vec{q}-\vec{p}) = \Bigl\{\gamma(\vec{q},r),\gamma^{\dagger
}(\vec{p},s)\Bigr\} \; .
\end{eqnarray}
The scalar wave function obeys a complicated equation,
\begin{equation}
\ddot{A}(t,k) + 3 H \dot{A}(t,k) + \frac{k^2}{a^2} A(t,k) + \xi (12 H^2 + 6 
\dot{H}) A(t,k) = 0 \; .
\end{equation}
It can be solved for any $a(t)$ in the conformally coupled case 
of $\xi = \frac16$,
\begin{equation}
{\rm any} \; a(t) \; , \xi = \frac16 \qquad \Longrightarrow \qquad 
A(t,k) = \frac1{\sqrt{2 k} \, a(t)} e^{- i k {\Delta \eta}(t)} \; .
\end{equation}
The general solution is also known for the minimally coupled case of $\xi = 
0$,\cite{TW3} but it is sufficiently complicated that I shall specialize to
the local de Sitter scale factor $a(t) = e^{Ht}$,
\begin{equation}
a(t) = e^{Ht} \; , \xi = 0 \qquad \Longrightarrow \qquad A(t,k) = 
\frac1{\sqrt{2 k}} \Bigl(\frac1{a(t)} + \frac{i H}{k}\Bigr) e^{- i k {\Delta 
\eta}(t)} \; .
\end{equation}
Of course the spinor wave functions are those of flat space expressed in 
terms of conformal time,
\begin{equation}
B(t,\vec{q},r) = \frac{u(\vec{q},r)}{\sqrt{2 q}} e^{-i q {\Delta \eta}(t)}
\qquad , \qquad C(t,\vec{p},s) = \frac{v(\vec{p},s)}{\sqrt{2 p}} 
e^{i p {\Delta \eta}(t)} \; .
\end{equation}

The time evolution operator of the interaction picture is,
\begin{equation}
U \equiv T\left\{ \exp\left[ -i f \! \int_{t_{\rm in}}^{t_{\rm out}} \!\!\!\!\!
\!\!\!\! dt \int \! d^3x \phi_I(t,\vec{x}) \overline{\Psi}_{\! I}(t,\vec{x}) 
\Psi_{\! I}(t,\vec{x}) \right] \right\} \; .
\end{equation}
We can think of the creation and annihilation operators as those relevant to
the initial time $t_{\rm in}$. The annihilators relevant to the final time
$t_{\rm out}$ are,
\begin{equation}
\alpha^{\rm out}\!(\vec{k}) = U^{\dagger} \alpha(\vec{k}) U \; , \;
\beta^{\rm out}\!(\vec{q},r) = U^{\dagger} \beta(\vec{q},r) U \; , \;
\gamma^{\rm out}\!(\vec{p},s) = U^{\dagger} \gamma(\vec{p},s) U .
\end{equation}
Now consider the amplitude for the initial vacuum to produce a scalar
and a fermion-anti-fermion pair,
\begin{eqnarray}
\lefteqn{\Bigl\langle \Omega \Bigl\vert \sqrt{2 k} \alpha(\vec{k}) \sqrt{2 q}
\beta(\vec{q},r) \sqrt{2 p} \gamma(\vec{p},s) U \Bigr\vert \Omega \Bigr\rangle}
\nonumber \\
& & = -i f (2\pi)^3 \delta^3(\vec{k} \!+\!\vec{q}\!+\! \vec{p}) \sqrt{8 k q p} 
\! \int_{t_{\rm in}}^{t_{\rm out}} \!\!\!\!\!\!\!\!\!\! dt A^*\!(t,k)
\overline{B}(t,\vec{q},r) C(t,\vec{p},s) + O(f^3) , \:\; \\
& & = -i f (2\pi)^3 \delta^3(\vec{k} \!+\!\vec{q}\!+\! \vec{p}) \,\overline{u}(
\vec{q},r) v(\vec{p},s) \!\int_{t_{\rm in}}^{t_{\rm out}} \!\!\!\!\!\!\!\!\!\! 
dt F(t) e^{i (k + q + p) {\Delta \eta}(t)} + O(f^3) . \:\: \label{ans}
\end{eqnarray}
The factor $F(t)$ in (\ref{ans}) is $1/a(t)$ for $\xi = \frac16$ and $(1/a(t)
\!-\! iH/k)$ for $\xi = 0$.

It is simple to understand why the amplitude for this process is zero in
flat space. For that case the scale factor is $a(t) = 1$, which implies the
conformal time interval is ${\Delta \eta}(t) = t$. One also usually takes
the initial and final times to $\pm \infty$, which results in a delta 
function that cannot be saturated for $\vec{k} + \vec{q} + \vec{p} = 0$,
\begin{equation}
{\rm Flat\ Space} \qquad \Longrightarrow \qquad \int_{\!-\!\infty}^{\infty}
\!\!\!\!\!\!\! dt e^{i (k + q + p) t} = 2 \pi \, \delta(k + q + p) \; .
\end{equation}
Even with a finite time interval $t_{\rm out}\!-\!t_{\rm in}$ the oscillations 
would still tend to cancel for intervals longer than $1/(k\!+\!q\!+\!p)$. This 
is the physics behind the energy-time uncertainty principle of flat space.

With $\xi = \frac16$ and arbitrary $a(t)$ the integral becomes,
\begin{equation}
{\rm Conformal\ Coupling} \qquad \Longrightarrow \qquad \int_{t_{\rm in}}^{
t_{\rm out}} \!\!\!\!\! \frac{dt}{a(t)} e^{i (k + q + p) \! \int_0^t \!
\frac{dt'}{a'}} \; .
\end{equation}
During inflation the phase factor approaches a constant, so there are no 
more oscillations. However, the integral is suppressed by the multiplicative
factor of $F(t) = 1/a(t)$. This is why inflation gives only a slight 
enhancement of quantum effects for massless, conformally invariant particles.

With minimal coupling in a locally de Sitter background the integral is,
\begin{equation}
{\rm Minimal\ Coupling} \quad \Longrightarrow \quad \int_{t_{\rm in}}^{
t_{\rm out}} \!\!\!\!\!\!\! dt \Bigl(e^{-\!H t} - \frac{iH}{k}\Bigr)
e^{i (k + q + p) (1 - e^{-H t})/H} \; .
\end{equation}
At late times (i.e., $H t \gg 1$) the integrand approaches a nonzero constant,
so the integral grows linearly in $t_{\rm out}$. Note also that a small mass 
$m \ll H$ would not begin to produce oscillations until a time comparable to 
$1/m$.

\section{Discussion}

We have just studied the mechanism through which massless, minimally coupled 
scalars catalyze the production of massless fermions during 
inflation.\cite{PW1} In a more complicated theory it is conceivable that 
this process might result in baryogenesis during inflation. A very similar 
calculation in massless, minimally coupled scalar QED gives strikingly
different results. In that case the one loop vacuum polarization causes
super-horizon photons to behave, in some ways, as if they possess nonzero 
mass.\cite{PTW1,PTW2,PW3} Although there is no significant creation of 
photons during inflation, their 0-point energies are vastly enhanced. After the
end of inflation some of this energy may end up seeding the cosmic magnetic 
fields we see in galaxies and galactic clusters.\cite{DPTD,PW1}

Massless, minimally coupled scalars with self-interactions can also do
interesting things to gravity through the back-reaction from the 
stress-energy tensor. What happens for a $\phi^4$ coupling depends upon 
the operator ordering scheme. If covariant normal-ordering is employed the 
three-loop back-reaction slows inflation by an amount that eventually becomes 
non-perturbatively strong.\cite{AW} The mechanism seems to be that inflation
rips virtual scalars out of the vacuum, whereupon the attractive, long range
interaction between these particles tends to pull them back together.

If time-ordering is used instead, the two-loop back-reaction increases the 
expansion rate.\cite{OW} The mechanism in this case seems to be that, as
more and more virtual scalars are ripped out of the vacuum, the amplitude
of the scalar field increases. This increases the potential energy from
the $\phi^4$ term, hence the expansion rate also increases. Starobinski\u{\i}
and Yokoyama have shown that the effect is finally arrested by the classical
force pushing the scalar towards the minimum of its potential.\cite{SY}

Although scalar effects are very interesting, they tend to be self limiting
because secular contributions to the scalar mass also occur. This is
guaranteed not to happen for gravitons, although the calculations are much
more difficult. The two-loop back-reaction from graviton creation slows 
inflation by an amount that eventually becomes non-perturbatively 
strong.\cite{TW1,TW2} The mechanism seems to be the same as for covariantly
normal-ordered scalars. It is conceivable that this process could quench 
$\Lambda$-driven inflation without the need for a scalar inflaton.

\section*{Acknowledgments}
It is a pleasure to acknowledge conversations and collaboration with
L. R. Abramo, V. K. Onemli, T. Prokopec and N. C. Tsamis. This work 
was partially supported by DOE contract DE-FG02-97ER41029 and by the 
Institute for Fundamental Theory at the University of Florida.

\end{document}